\documentclass[twoside]{article}
\usepackage[latin1]{inputenc}
\usepackage[OT1]{fontenc}
\usepackage{actes}

\ifx\WhizzyStart\undefined\relax\else\textheight 15cm\fi

\def\NotreTitre{Typer la dé-sérialisation sans sérialiser les types}
\usepackage[english,francais]{babel}
\usepackage{frbib}

\title{\NotreTitre}
\author{Grégoire {\sc Henry}$^1$, Michel {\sc Mauny}$^2$ et Emmanuel {\sc Chailloux}$^3$}
\titlehead{\NotreTitre}
\authorhead{Henry, Mauny \& Chailloux}
\affiliation{\begin{tabular}{rr} 
    \\ 1:  Laboratoire Preuves, Programmation et Systèmes (PPS),
    \\ CNRS UMR 7126, Université Denis Diderot, Paris
    \\ {\tt Gregoire.Henry@pps.jussieu.fr}
    \\ 2:  ENSTA, 32 Boulevard Victor, 75739 Paris cedex 15
    \\     INRIA-Rocquencourt, Domaine de Voluceau, 78153 Le Chesnay
    Cedex
    \\ {\tt Michel.Mauny@\{ensta,inria\}.fr}
    \\ 3:  Laboratoire Preuves, Programmation et Systèmes (PPS),
    \\ CNRS UMR 7126, Université Pierre et Marie Curie, Paris
    \\ {\tt Emmanuel.Chailloux@pps.jussieu.fr}
\end{tabular}}

\newtheorem{Lemme}{Lemme}

\newtheorem{Theoreme}{Théorème}
\newenvironment{Preuve}{\par\noindent{\bf
    Preuve}\relax}{\par}

\newenvironment{SPreuve}{\par\noindent{\bf
    Schéma de preuve}\relax}{\par}

\usepackage{mathpartir}
\usepackage{widearray}
\usepackage{ocamlweb}
\usepackage{latexsym}
\usepackage{amssymb}
\usepackage{color}
\usepackage{url}

\usepackage{guill}

\def\ML{ML}
\def\ocaml{{OCaml}}

\def\valml{Val-ML}

\def\cfc{CFC}
\def\ocamlcode#1{{\begin{otherlanguage}{english}\tt
      #1\end{otherlanguage}}}
\def\Fr#1{\begin{otherlanguage}{francais}#1\end{otherlanguage}}
\def\trepr#1{\hbox{\Fr{\tt«}}#1\hbox{\Fr{\tt»}}}

\def\ang#1{{\em #1}}

\def\Metarule#1{\vdash^{\hskip-1ex\hbox{\scriptsize${}^{{}_#1}$}}}

\def\kwd#1{\textrm{\bfseries #1}}
\def\comment#1{\hfill \mbox{\it #1}}%

\def\unn#1{{#1}_1,\ldots,{#1}_n}

\def\fv#1{\textrm{\sl fv}({#1})}

\def\e{e}
\def\f{e'}
\def\punn{\unn{p}}
\def\runn{\unn{r}}
\def\eunn{\unn{\e}}
\def\runn{\unn{r}}
\def\vunn{\unn{\v}}
\def\C#1{{\sf{C}}_{#1}}
\def\F#1#2{{\sf{F}}_{#1}(#2)}
\def\Let#1#2#3{\kwd{let}~(#1) = #2~\kwd{in}~#3}
\def\Letz{\kwd{let}}
\def\Fix#1#2{\kwd{fix}~(#1) = #2}
\def\Fixz{\kwd{fix}}

\def\TypeName#1{\mathsf{#1}}

\def\TInt{\TypeName{Int}}
\def\TConstrz#1{{#1}}
\def\TConstr#1#2{{#1}({#2})}

\def\tauunn{\unn{\tau}}

\def\Tbty{\textrm{\sf [}\;}
\def\Tor{\;\textrm{\sf l}\;}
\def\Tety{\;\textrm{\sf ]}}
\def\TRulez{\Metarule{{M\!L}}}
\def\TRule#1#2#3{ {#1} \TRulez {#2}:{#3}}
\def\TrecRulez{\Metarule{{M\!L}\!r\!e\!c}}
\def\TrecRule#1#2#3{ {#1} \TrecRulez {#2}:{#3}}

\def\vector#1{\overrightarrow{{#1}}}
\def\dom#1{\textrm{\sl dom}({#1})}
\def\domr#1{\textrm{\sl dom}({#1})}
\def\domp#1{\textrm{\sl dom}({#1})}
\def\img#1{\textrm{\sl img}({#1})}
\def\Tgen#1#2{\textrm{\sl gen}({#1},{#2})}

\def\TEq{=_{\alpha}}
\let\inst\leq
\let\geninst\preceq
\let\geninstr\succeq

\def\VKwd#1{\mathsf{#1}}
\def\v{v}
\def\w{w}
\def\vunn{\unn{\v}}
\def\VBlock#1#2{\VKwd{Block}(#1,#2)}
\def\VLet#1#2#3{\VKwd{let}~(#1)=#2~\VKwd{in}~#3}
\def\VFix#1#2{\VKwd{fix}~(#1) = #2}

\def\serial#1{{[\![\,#1\,]\!]}}

\def\UTypeName#1{\mathsf{#1}}
\def\UFunName#1{\textit{{#1}}}
\def\univ#1{\textrm{\sl{univ\/}}({#1})}
\def\univz{\textrm{\sl{univ\/}}}
\def\utau{\bar{\sigma}}
\def\utauunn{\unn{\utau}}

\def\UGamma{\bar{\Gamma}}
\def\UInt{\UTypeName{Int}}
\def\UTop{\top}
\def\UBottom{\bot}

\let\antiunif\wedge
\def\URulez{\Metarule{\top}}
\def\URule#1#2#3{ {#1} \URulez {#2}:{#3}}

\def\Tschema#1{\UFunName{schema}({#1})}

\def\check#1#2#3{\textrm{\sl Check}({#1},{#2},{#3})}
\def\checkz{\textrm{\sl Check}}

\def\KWD#1{\textrm{\bfseries #1}}
\def\LET{\KWD{let}}

\def\IN{\KWD{in}}
\def\IF{\KWD{if}}
\def\THEN{\KWD{then}}
\def\ELSE{\KWD{else}}
\def\FAIL{\KWD{Failure}}

\begin{document}
  \setcounter{page}{1}
  \maketitle

\section{Introduction}

La sérialisation\footnote{En Anglais, on dit {\em serialization} ou
{\em marshalling}, pour une mise en rang comme le ferait un officier
({\em marshall} ou maréchal), ou encore {\em pickling}, qui relève de
la mise en conserve, comme en témoignent les {\em pickles}, petits
cornichons conservés dans du vinaigre. {\em Marshalling} et {\em
marshall}, s'écrivent indifféremment avec un ou deux
$\ell$~\cite{webster}.} d'une valeur consiste à la représenter sous la forme d'une
suite d'octets de sorte à pouvoir la sauvegarder dans un fichier pour
relecture ultérieure ou la communiquer à d'autres programmes. Les
langages de programmation statiquement typés cherchent à fournir, lors
de la manipulation de ces valeurs dé-sérialisées, les mêmes garanties
de sûreté que pour les autres valeurs. C'est pourquoi on adjoint
généralement aux valeurs sérialisées une information supplémentaire
permettant d'attribuer un type correct à ces valeurs lors de leur
dé-sérialisation (voir par exemple
\cite{HerlihyLiskov1982,leroy93dynamics,JavaSerialization,jfpwjfla2000,Leifer2003}).

Le désavantage de l'adjonction de cette information de type est
qu'elle peut rendre difficile la transmission de valeurs entre
programmes écrits dans des langages différents, ou bien entre
différentes versions d'un même programme. Le choix effectué par
Objective Caml~\cite{ocaml} de ne pas intégrer d'informations de type
autres que celles nécessaires à la reconstruction de la valeur en
mémoire dans les valeurs sérialisées est un facteur de simplicité,
d'efficacité et de compacité. En conséquence, les fonctions de
dé-sérialisation d'\ocaml{} ne donnent lieu à aucune vérification, et
la sécurité du typage repose sur la qualité des informations de type
associées par le programmeur aux valeurs dé-sérialisées. Le programme
utilisant ces fonctions de dé-sérialisation court donc le risque de
rencontrer une erreur de type à l'exécution.

Dans cet article, nous remédions à ce problème en proposant une façon
de donner un type statique aux fonctions de dé-sérialisation en \ocaml{},
d'utiliser ces types statiques pour effectuer des vérifications
dynamiques lors de la dé-sérialisation et nous prouvons la sûreté de
ces mécanismes. À l'évidence, lorsque les valeurs considérées sont ou
bien des données atomiques ou alors des données structurées (non
fonctionnelles) sans partage ni cycle, la vérification de
l'appartenance d'une telle valeur à un type donné est très facile à
mettre en {\oe}uvre: un simple parcours récursif est suffisant. Par
contre, le problème devient plus complexe lorsqu'on est en présence de
partage ou de cycles, sources potentielles de polymorphisme.

Après une description des types que nous attribuons aux fonctions de
dé-sérialisation à la section~\ref{sec:reprTypes}, nous décrivons
informellement l'ensemble du processus de vérification à la
section~\ref{sec:informel}. La section~\ref{sec:progVal} décrit quels
sont les programmes et les valeurs que nous considérons, ainsi que la
traduction des programmes en valeurs. La section~\ref{sec:check}
décrit l'algorithme de vérification, et la section~\ref{sec:props} en
énonce les propriétés, et indique quelles sont les techniques
utilisées pour les prouver. La section~\ref{sec:impl} décrit
brièvement les implantations prototypes qui ont été réalisées, et les
sections~\ref{sec:connexes} et \ref{sec:futur} mentionnent des travaux
connexes et discutent quelques axes de travail futurs.


\section{Représentation dynamique des types statiques}\label{sec:reprTypes}

Vérifier dynamiquement le type d'une valeur dé-sérialisée impose
d'avoir accès à une représentation de ce type au moment de la
dé-sérialisation. Si c'était le cas, les fonctions de
dé-sérialisation de valeurs depuis une chaîne de caractères ou un
fichier pourraient avoir les types\footnote{Nous suivons dans cet
article la convention de notation préfixe des constructeurs de types
où les noms de ces constructeurs commencent par une majuscule. Par
exemple, le type \ocaml{} \ocamlcode{int list} est écrit ici
\ocamlcode{List(Int)}.}  suivants:
\begin{quote}
\ocamlcode{Marshal.fromString : $\forall \alpha \,.\,$TRepr($\alpha$) $\to$
String $\to$ $\alpha$}\\
\ocamlcode{Marshal.fromFile : $\forall \alpha \,.\,$TRepr($\alpha$) $\to$
Filename $\to$ $\alpha$}
\end{quote}
où \ocamlcode{TRepr} est un type paramétré tel que la seule
valeur du type \ocamlcode{TRepr($\tau$)} est une description de la
représentation des valeurs du type $\tau$. Pour $\tau$ donné, nous
noterons \ocamlcode{\trepr{$\tau$}} cette valeur de type
\ocamlcode{TRepr($\tau$)} dans les exemples de programmes considérés
dans la suite de cet article. Le programme suivant illustre l'usage
d'une fonction de dé-sérialisation typée de cette façon:
\begin{quote}
\ocamlcode{let xs = Marshal.fromFile \trepr{List(Int)} "/tmp/f.dat" \\
  in List.foldLeft (+) 0 xs}
\end{quote}
Afin d'éviter toute interaction entre les variables de types
pouvant apparaître dans \ocamlcode{\trepr{$\tau$}} et dans le reste du
programme, nous considérons qu'implicitement cette notation quantifie
universellement les variables apparaissant dans $\tau$ et représente
ainsi un schéma de type clos.

Cette possibilité de disposer des représentations de types à
l'exécution sous forme de types singletons a été proposée par Crary,
Weirich et Morrisett dans~\cite{crary98intensional} et utilisée comme
nous le faisons ici par Hicks, Weirich et Crary dans~\cite{HicksTIC2000}.

\section{Présentation informelle}\label{sec:informel}

Avant de procéder à la présentation détaillée des langages et systèmes
de types nécessaires à la description de l'algorithme et à
l'énoncé de ses propriétés, nous donnons dans cette section une
description informelle du processus de vérification des valeurs
dé-sérialisées.

\paragraph{Parcourir type et valeur en parallèle}
Les fonctions de dé-sérialisation, munies d'une représentation de
schéma de type, ont à vérifier que la représentation mémoire de la
valeur $v$ qui leur est présentée est compatible avec ce schéma de
type. Cette vérification, qui utilise le type comme une contrainte que
la valeur doit satisfaire, est réalisée par un parcours complet de la
valeur avec, en parallèle, une expansion progressive du corps du
schéma de type de sorte à exhiber le corps des définitions de
constructeurs de type. Cette expansion permet, par exemple, de savoir
que les seules valeurs possibles du type \ocamlcode{List($\tau$)} sont
représentées ou bien par un entier\footnote{Le numéro d'ordre de
constructeur constant dans la liste des constructeurs constants du
type.}  (pour le constructeur \ocamlcode{[$\,$]}) ou alors par un bloc
marqué à 2 champs contenant respectivement des valeurs de types
\ocamlcode{$\tau$} et \ocamlcode{List($\tau$)}. Notons que la
représentation mémoire d'une valeur donnée peut être compatible avec
plusieurs types non reliés: par exemple, l'entier 0 peut être
indifféremment vu comme booléen, entier, ou comme le premier
constructeur constant d'un type arbitraire.

\paragraph{Isomorphismes de types et polymorphisme} Si,
durant ce parcours, nous sommes amenés à vérifier que la valeur est de
type \ocamlcode{$ \forall \alpha\,.\,$(List($\alpha$)$\times$List($\alpha$))} et que la valeur courante est bien la représentation
d'une paire \ocamlcode{($v_1$, $v_2$)}, nous nous ramènerons à
vérifier que $v_1$ et $v_2$ sont toutes deux de type
\ocamlcode{$\forall\alpha\,.\,$List($\alpha$)}, c'est-à-dire
\ocamlcode{List($\forall\alpha\,.\,\alpha$)}. Intuitivement, une
valeur de type \ocamlcode{$ \forall \alpha\,.\,$(List($\alpha$)$\times$List($\alpha$))} est nécessairement une paire de listes vides, et est
donc aussi de type \ocamlcode{(List($ \forall \alpha\,.\,\alpha$)$\times$List($ \forall \alpha\,.\,\alpha$))}. Il s'agit là d'isomorphismes de
types~\cite{DBLP:journals/jfp/Cosmo93} qui nous indiquent que
\ocamlcode{$ \forall \alpha\,.\,$(List($\alpha$)$\times$List($\alpha$))} est
équivalent à \ocamlcode{$ \forall \alpha\beta\,.\,$(List($\alpha$)$\times$List($\beta$))}, qui lui-même est équivalent à \ocamlcode{(List($
\forall \alpha\,.\,\alpha$)$\times$List($ \forall
\beta\,.\,\beta$))}. Le polymorphisme des
valeurs de ce langage signifie une absence de composants (d'éléments,
pour une liste, par exemple). Cette remarque permet donc de renommer
les variables de type de sorte que chaque occurrence de variable soit
distincte de toutes les autres, et invite à rapprocher chaque
quantificateur de la variable qu'il quantifie. On transforme donc
naturellement le type $\forall\unn{\alpha}.\tau$ en
$\tau[(\forall\beta.\beta)/\alpha_i]_{i=1..n}$. L'implication
essentielle de cette remarque et de cette transformation sur
l'algorithme de vérification est que si on est amené à vérifier qu'une
(partie de) valeur est de type $\forall\alpha.\alpha$, alors
l'algorithme termine en échec\footnote{La vacuité du type
$\forall\alpha.\alpha$ est motivée par le fait que nous ne considérons
que des données structurées complètement évaluées et non pas des
calculs, comme on pourrait le faire dans un langage paresseux.}.

Au lieu d'effectuer cette transformation, nous introduisons une
nouvelle constante notée $\UTop$, qui représentera
$\forall\alpha.\alpha$, et on traduira le schéma
$\forall\unn{\alpha}.\tau$ en le type $\tau[\UTop/\alpha_i]_{i=1..n}$,
obtenu en remplaçant toutes les variables de type apparaissant dans
$\tau$ par $\UTop$. Nous noterons $\utau$ les types obtenus de cette
façon. L'algorithme travaillera donc avec un terme sans variables au
lieu d'un schéma de type, et devra échouer lorsqu'il aura à vérifier que
la valeur courante est du type $\UTop$.
L'équivalence entre ces deux notions de types 
sera formalisée par le théorème \ref{theoreme:equivalence-u-ml}.

\paragraph{Partage} Si la valeur à vérifier est un arbre (sans
partage, donc), alors la vérification consiste en un simple parcours
de la valeur en profondeur d'abord. Si, par contre, certains blocs
sont partagés, il est tout-à-fait possible que l'un des pointeurs vers
ce bloc contraigne ce dernier à être de type $\utau_1$ et qu'un autre
pointeur lui impose d'être de type $\utau_2$. Si $\utau_1$ et
$\utau_2$ sont incompatibles, une façon pour ce bloc d'être à la fois
de type $\utau_1$ et $\utau_2$ est d'être d'un type que nous noterons
$\utau_1 \antiunif \utau_2$ et qui est un
anti-unificateur~\cite{plotkin-phd, huet-these} de $\utau_1$ et
$\utau_2$: intuitivement, pour être à la fois de type $\utau_1$ et
$\utau_2$, la valeur ne doit pas avoir de composante qui soit
particulière à un seul de $\utau_1$ ou $\utau_2$. Notons ici qu'une
fois cet anti-unificateur calculé, nous pouvons ne parcourir qu'une
seule fois ce bloc partagé et les valeurs qui en sont issues.

\paragraph{Cycles} En présence d'un cycle, nous
devons distinguer les pointeurs en provenance de l'«intérieur du
cycle», c'est-à-dire de la composante fortement connexe (\cfc{}) à
laquelle appartient ce cycle, des pointeurs en provenance de
l'«extérieur» de la \cfc{}, c'est-à-dire du {\em contexte} de cette
\cfc{}. En effet, c'est le contexte qui fixera le type demandé à la
CFC: les pointeurs en provenance du contexte sont porteurs de types
qui sont anti-unifiés, comme précédemment.  Une fois connues les
contraintes de type imposées à chacun des n{\oe}uds racines de la CFC
par son contexte, on parcourt récursivement cette CFC à partir
d'une de ces racines, muni de la contrainte de type associée. Durant ce
parcours, lorsqu'on rencontre des pointeurs vers les n{\oe}uds racines
de la CFC, on vérifie que la contrainte associée à chacun de ces
pointeurs est une instance de la contrainte émise par le contexte qui
pèse sur ce n{\oe}ud.

\bigskip

Pour résumer cette présentation informelle de l'algorithme, le
parcours de la valeur va être effectué en profondeur d'abord,
retardant la vérification des n{\oe}uds partagés. Pour une valeur $v$
dont le n{\oe}ud racine est partagé $n$ fois, une fois atteints les
$n$ pointeurs qui le désignent, on parcourt la valeur $v$
récursivement. Si on a parcouru entièrement le contexte de $v$ en ne
rencontrant que $m<n$ pointeurs vers $v$, alors le n{\oe}ud racine de
$v$ fait partie d'un cycle: l'algorithme calcule alors ses CFC et les
traite récursivement, l'un après l'autre, dans l'ordre topologique de
dépendance.

\section{Programmes et valeurs}\label{sec:progVal}

Avant de décrire en détail la vérification  des valeurs
dé-sérialisées, fixons plus précisément les limites de cette étude:
nous ne considérons dans cet article que les valeurs constituées de
types de base et de types algébriques (enregistrements et unions
discriminées). En particulier, on ne considèrera pas ici les objets et
valeurs fonctionnelles, les laissant comme
axes de travail futur: ils posent en effet des problèmes {\em a
priori} plus difficiles que les valeurs que nous étudions. Nous ne
considérons pas non plus pour le moment les types de données mutables: nous y
reviendrons à la fin de cet article.

L'objectif principal est bien sûr de prouver que l'algorithme de
vérification est correct (lemme \ref{lemme:check-correction-induction}
et théorème \ref{theoreme:check-correction}), c'est-à-dire que si la
vérification de $v:\utau$ réussit, alors un programme $e$ dont la
sérialisation serait $v$ possède effectivement tous les types
représentés par $\utau$. Cela implique, par le théorème de correction
du typage des programmes, que cette valeur $v$ peut être utilisée dans
un contexte d'exécution attendant une valeur de ce type.

Un autre objectif est la complétude de l'algorithme (lemme
\ref{lemme:check-complete-induction} et théorème
\ref{theoreme:check-complete}), c'est-à-dire que tout programme $e$ de
type $\tau$ peut être sérialisé en une valeur $v$ dont la vérification
par l'algorithme réussira. Malheureusement, cette propriété n'est
satisfaite que si la preuve de typage de $e:\tau$ n'utilise pas la
règle de typage polymorphe de la
récursion~\cite{MycroftRec,kfoury-polyrec}. Dans la pratique, l'effet
de cette limitation restera probablement très marginal.

\subsection{Programmes}

Ce langage de valeurs correspond à un sous-ensemble des programmes
source possibles. Puisque nous nous situons dans le cadre des langages
à la ML, le langage source considéré, que nous appelons
\valml{}, est donné par la grammaire suivante:
\begin{widearray}{LRZ}
\e & ::= & n \comment{entier}\\
   &   | & p \comment{variable indiquant un partage} \\
   &   | & r \comment{variable indiquant un cycle} \\
   &   | & (\e,\ldots,\e) \comment{n-uplet} \\
   &   | & \C{i}\comment{constructeur constant}\\
   &   | & \F{i}{\e} \comment{constructeur non constant} \\
   &   | & \Let{\punn}{\f}{\e} \qquad\hbox{\rm avec~} p_i \in \fv{\e}, \forall i=1..n
           \comment{partage} \\
   &   | & \Fix{\runn}{e} \qquad\qquad\hbox{\rm avec~}  r_i \in
   \fv{\e}, \forall i=1..n
           \comment{cycle}\\
\end{widearray}
où $\fv{\e}$ désigne l'ensemble des variables libres d'un programme
$e$.

Notons que l'on impose aux déclarations $\Letz$ et $\Fixz$ des
programmes d'être {\em utiles}, c'est-à-dire de n'introduire que des
noms de variables qui sont effectivement utilisés dans leur portée. La
raison en est que ces programmes seront traduits en des valeurs qui
doivent être des graphes connexes. Une déclaration de variable
inutilisée contribuerait à produire un graphe non connexe.

De plus, on prend soin de distinguer les pointeurs «internes» à une CFC
notés $r$ et introduits par la construction récursive $\Fixz$,
des pointeurs «externes» symbolisant simplement le partage, notés $p$, et introduits
par la construction $\Letz$.

\paragraph{Types des programmes}

Les expressions de type $\tau$ sont formées à partir de types de base, de
variables de types, et de types paramétrés. La notion de schéma de
type (notés $\sigma$) est classique, ainsi que les définitions de
type, notées $\delta$.
\begin{widearray}{LRZ}
  \tau & ::= & \alpha  
       ~   | ~ \TInt  
       ~   | ~ (\tau \times \ldots \times\tau)\
       ~   | ~ \TConstr{T}{\vector\tau}\\
\sigma & ::= & \forall \alpha . \tau ~|~ \tau\\
\delta & ::= & \TConstr{T}{\vector\alpha}
            =  \Tbty\C{1}\Tor\ldots\Tor\C{n}\Tor
            \F{1}{\tau}
            \Tor\ldots
            \Tor\F{k}{\tau}\Tety
\end{widearray}
Les types de n-uplets sont notés comme des produits cartésiens
d'arité variable.
Pour simplifier la présentation, ces définitions de types sont
considérées comme mutuellement récursives, et les constructeurs de
données qu'elles introduisent sont supposés distincts les uns des
autres, afin qu'un constructeur de données identifie de façon
non-ambiguë un constructeur de type. Ici, $\vector\alpha$ représente
les paramètres formels du constructeur de types $T$, les $\C{i}$ sont
ses constructeurs constants et les $\F{i}{\_}$ ses constructeurs non
constants\footnote{Nous ne considérons ici que les constructeurs de
données constants ou unaires afin de simplifier la présentation. Les
constructeurs n-aires induisent en effet des cas de preuve
partiellement redondants avec le cas des n-uplets.}.

Pour un programme donné, l'ensemble de ses définitions de types est
noté $\Delta$.
En notant $\theta$ la substitution qui associe les $\vector\tau$ aux
$\vector\alpha$, on pose, pour $T(\vector\alpha)$ défini dans
$\Delta$:
\begin{widearray}{Z}
\Delta(\TConstr{T}{\vector\tau}) = 
  \Tbty \C{1}\Tor
  \ldots\Tor \C{n}\Tor \F{1}{\theta(\tau_1)}\Tor \ldots\Tor \F{k}{\theta(\tau_k)}\Tety\\
\end{widearray}
ce qui nous permet d'accéder directement à la définition instanciée de
$\TConstrz{T}$.

\paragraph{Environnement de typage}
Les environnements de typage sont notés $\Gamma$ et définis par:
\begin{widearray}{LRZ}
  \Gamma & ::= & \emptyset ~|~ \Gamma \oplus \{ p: \sigma \} ~|~ \Gamma
  \oplus \{ r: \sigma \}
\end{widearray}
On utilisera souvent $\Gamma$ comme une fonction associant à un nom de
pointeur $q$ (où $q$ est un pointeur $p$ ou $r$) un schéma de type
$\Gamma(q)$.  On notera $\dom{\Gamma}$ l'ensemble des variables $q$
pour lesquels $\Gamma(q)$ est défini et,
$\Gamma\setminus\{q_1,\ldots,q_n\}$ la restriction de $\Gamma$ à
$\dom{\Gamma}\setminus\{q_1,\ldots,q_n\}$. Enfin, on abrégera la
généralisation d'un type $\tau$ dans un environnement $\Gamma$ en notant:
\begin{widearray}{RCZ}
\Tgen{\Gamma}{\tau} &=& 
    \forall\vector{\fv{\tau}\backslash\fv{\Gamma}}\cdot\tau
\end{widearray}

\paragraph{Instanciation (polymorphe)}
On notera:
\begin{itemize}
\item $\tau \inst \sigma$ ($\tau$ est une instance polymorphe de
  $\sigma$), si $\sigma$ s'écrit
  $\forall\vector{\alpha}.\tau'$ et s'il existe une substitution
  $\theta$ telle que $\dom{\theta}=\{\vector{\alpha}\}$ et
  $\tau=\theta(\tau')$;
\item $\sigma \geninst \sigma'$ si toute instance polymorphe de
  $\sigma$ est aussi instance polymorphe de $\sigma'$ (en d'autres termes,
$\sigma'$ contient plus d'instances de type que $\sigma$).

\end{itemize}
Étendant la relation d'instanciation polymorphe aux
environnements de typage, on écrira $\Gamma \geninst \Gamma'$ si:
\begin{itemize}
\item $\dom{\Gamma} = \dom{\Gamma'}$ ;
\item $\forall p\in\domp{\Gamma},\;\Gamma(p) \geninst \Gamma'(p)$ ;
\item $\forall r\in\domr{\Gamma},\;\Gamma(r) \TEq
  \Gamma'(r)$.
\end{itemize}
Remarquons que les pointeurs récursifs reçoivent un traitement
spécifique: pour instancier un environnement, on ne peut instancier
que les types associés à des pointeurs non récursifs (notés $p$). Les
types associés à des pointeurs récursifs (notés $r$) doivent, quant à
eux, rester inchangés à un renommage près, comme de coutume. Cette
distinction est due au traitement différent reçu par les pointeurs de
partage --- pour lesquels des types distincts sont anti-unifiés --- de
celui reçu par les pointeurs récursifs, dont on vérifie simplement que
le type est une instance du type du cycle dont ils sont racine.

\subsection{Valeurs}

Le langage des valeurs que nous considérons peut être représenté par
la grammaire suivante. Les blocs représentent de la mémoire allouée et
sont dotés d'une marque entière $i$, d'une taille $n$ et
de $n$ valeurs contenues dans chacun des champs. Tous les pointeurs
sont déclarés: ils indiquent la présence de partage ou de cycles
effectifs. Aucun pointeur déclaré n'est superflu. La construction de
ces valeurs à partir de leur représentation sérialisée est décrite à la
section \ref{sec:impl}.

\begin{widearray}{LRZ}
\v & ::= & ~ n ~~|~~  p ~~|~~  r \comment{entier, pointeurs et pointeurs récursifs} \\
   &   | & \VBlock{i}{\vunn} \comment{bloc alloué de
            marque $i \geq 1$ et d'arité $n \geq 1$} \\
   &   | & \VLet{\punn}{\v}{\w} \comment{partage}\\
   &   | & \VFix{\runn}{\v} \comment{cycles (composantes fortement connexes)}
\end{widearray}

\paragraph{Types des valeurs}\label{sec:tyvaleurs}

Les types des valeurs sont construits comme les types des programmes
où une constante notée $\UTop$ représentant le type vide, aura pris la
place des variables de types.  On les notera comme des termes de la
grammaire suivante:
\begin{widearray}{LRZ}
  \utau & ::= & \UTop ~|~ \TInt ~|~ (\utau_1\times\ldots\times\utau_n) ~|~ \TConstr{T}{\utauunn}
\end{widearray}
On les définit formellement
comme les représentants de classes d'équivalence de schémas de types
clos. La
relation d'équivalence est notée $\sim$ et est définie par:
\begin{widearray}{LRZ}
\forall \alpha_1,\ldots,\alpha_n . \tau \sim \forall  \beta_1,\ldots,\beta_m . \tau'
\hbox{~si et seulement si~} \tau[\gamma/\alpha_1,\ldots,\alpha_n] = \tau'[\gamma/\beta_1,\ldots,\beta_m]
\end{widearray}
où $\gamma$ est une variable «fraîche». En d'autres termes, deux schémas sont équivalents s'ils ne diffèrent
qu'aux occurrences de variables. Il est immédiat que cette
relation est une relation d'équivalence.

On se convainc aisément que chacune de ces classes d'équivalence contient
des schémas de types qui sont isomorphes au sens que nous avons donné
à la section~\ref{sec:informel}, et le théorème \ref{theoreme:equivalence-u-ml}
démontre que, pour une expression close $e$ typable par un schéma de
type clos $\sigma$,  les différents membres de la classe  de $\sigma$
sont aussi des types valides pour $e$.

Le représentant de la classe d'un schéma $\sigma$ peut être
calculé par la fonction $\univz$ ainsi définie:
\begin{widearray}{RCZ}
\univ{\forall \alpha. \sigma} &=& \univ{\sigma}\\
\univ{\alpha} &=& \UTop\\
\univ{\TInt} &=& \UInt\\
\univ{(\tau_1\times\ldots\times\tau_n)} &=& 
          (\univ{\tau_1}\times\ldots\times\univ{\tau_n})\\
\univ{\TConstr{T}{\tauunn}} &=& 
          \TConstr{T}{\univ{\tau_1},\ldots,\univ{\tau_n}}\\
\end{widearray}
Inversement, on notera $\Tschema{\utau}$ le plus général des
schémas de type de la classe représenté par $\utau$.

On écrira $\utau \geninst \utau'$ si $\Tschema{\utau} \geninst
\Tschema{\utau'}$.  Et en notant $\UGamma$, $\UGamma'$, \ldots, les
environnements de typage composés d'assertions de la forme
$q:\utau$, on étend cette relation d'instanciation polymorphe aux
environnements de la même manière que pour les classes de types: en
exigeant l'égalité sur les pointeurs récursifs.


\subsection{Traduction des programmes en valeurs}

Le passage d'un programme à une valeur est assuré par la fonction de
traduction suivante:
\begin{widearray}{RCZ}
\serial{n} & = & n \\
\serial{p} & = & p \\
\serial{r} & = & r \\
\serial{(\eunn)} & = & \VBlock{0}{\serial{\e_1},\ldots,\serial{\e_n}} \\
\serial{\C{i}} & = & i \\
\serial{\F{i}{\e}} & = & \VBlock{i}{\serial{\e}} \\
\serial{\Let{\punn}{\f}{\e}} & = & \VLet{\punn}{\serial{\f}}{\serial{\e}}\\
\serial{\Fix{\runn}{\e}} & = & \VFix{\runn}{\,\serial{\e}\,}\\
\end{widearray}
Cette traduction identifie les entiers et les constructeurs constants de
même rang, ainsi que les constructeurs fonctionnels de même rang et de même
arité. La vérification de types à laquelle on procède ici a pour but
de lever les ambiguïtés introduites par cette traduction.

\subsection{Anti-unification sans mémoire}
L'anti-unification classique de deux termes du premier ordre produit
un troisième terme, appelé anti-unificateur, dont les deux premiers
sont des instances. Parmi les anti-unificateurs de deux termes donnés,
le meilleur est celui qui est le moins général, instance de tous les
autres. Classiquement, l'anti-unification produit donc des termes avec
variables. Nous en donnons ici une définition légèrement différente,
que nous appelons {\em unification sans mémoire} et qui n'utilise pas
de variables mais la constante $\UTop$, produite lorsque deux termes à
anti-unifier entrent en conflit. Là où l'anti-unification classique
produirait plusieurs fois la même variable, mémorisant ainsi plusieurs
occurrences d'un même conflit, nous produisons plusieurs occurrences
du terme $\UTop$ qui pourront être instanciées indépendamment les unes
des autres, conformément à l'intuition que nous procure les
isomorphismes de types mentionnés à la
section~\ref{sec:informel}.  L'anti-unification
sans mémoire de $\utau_1$ et $\utau_2$ est notée $\utau_1 \antiunif
\utau_2$, et est définie par:
\begin{widearray}{RCZ}
\TInt \antiunif \TInt &=& \TInt\\
(\utau_1\times\ldots\times\utau_n) \antiunif (\utau'_1\times\ldots\times\utau'_n) & = &
          (\utau_1 \antiunif \utau'_1\times\ldots\times\utau_n \antiunif \utau'_n)\\
\TConstr{T}{\utauunn} \antiunif \TConstr{T'}{\unn{\utau'}} &=&
     \TConstr{T}{\utau_1 \antiunif \utau'_1, \ldots, \utau_n \antiunif \utau'_n}
\comment{si $\TConstrz{T} = \TConstrz{T'}$}\\
\utau_1 \antiunif \utau_2 &=& \UTop \comment{dans tous les autres cas}
\end{widearray}
Dans la suite, ne faisant plus référence à l'anti-unification
classique, nous écrirons simplement {\em anti-unification} pour
dénoter l'anti-unification sans mémoire.

On ajoute à notre algèbre de types un élément noté $\UBottom$, neutre
pour l'anti-unification. Cet élément sera utilisé comme hypothèse de
type initiale pour les pointeurs lorsque l'algorithme entre dans leur
portée, et sera nécessairement anti-unifié avec le type d'au moins une
occurrence de ce pointeur: en effet, les contraintes de formation des
programmes, qui imposent aux variables liées d'apparaître libres au
noins une fois dans leur portée (aucune déclaration n'est inutile)
sont, par traduction, transférées aux valeurs. Il en résulte que
$\UBottom$ ne sera jamais utilisé comme argument de la fonction de
vérification.

\section{$\checkz$, algorithme de vérification de types}\label{sec:check}

La vérification est représentée par le calcul de
$\check{\UGamma}{\utau}{\v}$ où $\UGamma$ est un environnement, et
$\utau$ et $\v$ sont respectivement le type et la valeur à
vérifier. Un appel $\check{\UGamma}{\utau}{\v}$ produira un couple
$(e, \UGamma')$ ou bien échouera. En cas de succès, $\UGamma'$ est une
généralisation de $\UGamma$ (ce que montrera le lemme
\ref{lemme:check-correction-gamma-structure}), et $e$ est un programme
dont la représentation est la valeur $v$. Un appel initial à $\checkz$
est de la forme $\check{\emptyset}{\utau}{\v}$ et produira une paire
$(e, \emptyset)$ en cas de succès. Un invariant de cette fonction est
que $\fv{\v} \subseteq \dom{\UGamma}$: en d'autres termes tous les
pointeurs libres de $v$ se voient attribuer un type (éventuellement
$\UBottom$) dans $\UGamma$. Il va de soi que dans une
implémentation effective, cet algorithme se bornera à parcourir la
valeur examinée et la rendre en résultat si la vérification
réussit.


On notera $\UGamma \otimes \{p: \utau\} $, l'environnement $\UGamma'$
tel que
$\UGamma'(p) = \UGamma(p) \antiunif \utau$ et
$\UGamma'(p') = \UGamma(p')$ pour $p' \neq p$.


\begin{widearray}{Z}
\check{\UGamma}{\TInt}{i} = (i,\UGamma) 
\end{widearray}
%
%
\begin{widearray}{Z} 
\check{\UGamma}{\TConstr{T}{\utau_1,\ldots,\utau_m}}{i} = 
\\
\quad\LET~\Tbty\C{1}\Tor\ldots\Tor\C{n}
                \Tor\F{1}{\utau'_1}
                \Tor\ldots
                \Tor\F{k}{\utau'_k}\Tety~=\Delta(\TConstr{T}{\utau_1,\ldots,\utau_m})~\IN\\
\quad \IF~ 0 < i \leq n~\THEN~(\C{i}, \UGamma)~\ELSE~\FAIL\\
\end{widearray} 
%
%
\begin{widearray}{Z}
\check{\UGamma}{\TConstr{T}{\utau_1,\ldots,\utau_m}}{\VBlock{i}{\v'}} =
\\
\quad\LET~\Tbty\C{1}\Tor\ldots\Tor\C{n}
                \Tor\F{1}{\utau'_1}
                \Tor\ldots
                \Tor\F{k}{\utau'_k}\Tety~=\Delta(\TConstr{T}{\utau_1,\ldots,\utau_m})~\IN\\
\quad \IF~ 0 < i \leq k~\THEN\\
\quad\quad\LET~(\e',\UGamma')~=\check{\UGamma}{\utau'_i}{\v'}~\IN\\
\quad\quad (\F{i}{\e'}, \UGamma')\\
\quad\ELSE\\
\quad\quad\FAIL\\
\end{widearray}
%
%
\begin{widearray}{Z}
\check{\UGamma}{(\utau_1\times\ldots\times\utau_n)}{\VBlock{0}{\vunn}} =
\\
\quad\LET~(\e_1,\UGamma_1) = \check{\UGamma}{\utau_1}{\v_1}~\IN\\
\quad\ldots\\
\quad\LET~(\e_n,\UGamma_n) = \check{\UGamma_{n-1}}{\utau_n}{\v_n}~\IN\\
\quad((\eunn),\UGamma_n)\\
\end{widearray}
%
%
\begin{widearray}{Z}
\check{\UGamma}{\utau}{p}=
\\
\quad\quad (p,(\UGamma\otimes\{p:\utau\}))
\end{widearray}
%
%
\begin{widearray}{Z}
\check{\UGamma}{\utau}{r}=
\\
\quad\quad \IF~\utau \geninst \UGamma(r)~\THEN~(r,\UGamma)~\ELSE~\FAIL\\
\end{widearray}
%
%
\begin{widearray}{Z}
\check{\UGamma}{\utau}{(\VLet{\punn}{\v}{\w})} =
\\
\quad\LET~ (\e,\UGamma') = \check{\UGamma\oplus\{p_i:\UBottom\}_{i=1..n}}{\utau}{\w}~\IN \\
\quad\LET~ (\e',\UGamma'')=\check{\UGamma'\backslash\{\punn\}}{(\UGamma'(p_1),\ldots,\UGamma'(p_n))}{\v}~\IN\\
\quad((\Let{\punn}{\e'}{\e}),\UGamma'')
\end{widearray}
%
%
\begin{widearray}{Z}
\check{\UGamma}{(\utauunn)}{(\VFix{\runn}{\v})} =
\\
\quad\LET~\UGamma'=\UGamma \oplus\{ r_i: \utau_i\}_{i=1..n}~\IN\\
\quad\LET~(\e,\UGamma'')=\check{\UGamma'}{(\utauunn)}{\v}~\IN\\
\quad((\Fix{\runn}{\e}),\UGamma''\setminus \{ r_i \}_{i=1..n})\\
\end{widearray}
Dans tous les autres cas:
\begin{widearray}{Z}
\check{\UGamma}{\utau}{\v} = \FAIL
\end{widearray}

\section{Propriétés de l'algorithme}\label{sec:props}

Nous énonçons dans cette section les propriétés que satisfait
l'algorithme de vérification. Nous en donnons aussi les schémas de
preuves, mais invitons le lecteur à se reporter
à~\cite{henryMPRI}, où il pourra trouver les preuves
complètes.

\subsection{Typage des programmes}

La discipline de typage à laquelle sont soumis les programmes est la
discipline habituelle de typage de {\ML} avec récursion
polymorphe~\cite{MycroftRec}. Puisque nous ne considérons pas les
valeurs mutables, la généralisation n'est pas soumise à restriction.

\subsubsection{$\TrecRulez$, typage classique des programmes, avec récursion
  polymorphe}\label{sec:mlrec}

\begin{mathpar}
\inferrule{}{\TrecRule{\Gamma}{n}{\TInt}}
\and
\inferrule{\tau \inst \Gamma(p)}{\TrecRule{\Gamma}{p}{\tau}}
\and
\inferrule{\tau \inst \Gamma(r)}{\TrecRule{\Gamma}{r}{\tau}}
\and
\inferrule{ \forall i=1..n, ~ \TrecRule{\Gamma}{\e_i }{\tau_i}}
          {\TrecRule{\Gamma}{(\e_1, \ldots, \e_n) }{(\tau_1\times\ldots\times\tau_n)}}
\and
\inferrule{\C{i}\in\Delta(\TConstr{T}{\vector\tau})}
          {\TrecRule{\Gamma}{\C{i}}{T(\vector\tau)}}
\and
\inferrule{\F{i}{\tau'}\in\Delta(\TConstr{T}{\vector\tau})
  \\ \TrecRule{\Gamma}{\e}{\tau'}}
          {\TrecRule{\Gamma}{\F{i}{\e}}{T(\vector\tau)}}
\and
\inferrule{\TrecRule{\Gamma}{e' }{(\tau_1\times\ldots\times\tau_n)} \\ 
    \TrecRule{\Gamma \oplus \{ p_i: \Tgen{\Gamma}{\tau_i}
 \}_{i=1..n}}{\e}{\tau}}
          {\TrecRule{\Gamma}{\Let{\punn}{\e'}{\e}}{\tau}}
\and
\inferrule{ 
    \TrecRule{\Gamma \oplus \{ r_i: \Tgen{\Gamma}{\tau'_i} \}_{i=1..n}}{\e}{(\tau'_1\times\ldots\times\tau'_n)}}
          {\TrecRule{\Gamma}{\Fix{\runn}{\e}}{(\tau_1\times\ldots\times\tau_n)}}
\quad (\forall i=1..n,\; \tau_i \inst \Tgen{\Gamma}{\tau'_i})
\end{mathpar}

Rappelons \cite{kfoury-polyrec} que si $\Gamma' \geninstr \Gamma$ et 
\inferrule{}{\TrecRule{\Gamma}{e }{\tau}} alors 
\inferrule{}{\TrecRule{\Gamma'}{e }{\tau}}.

\subsubsection{$\URulez$, un système de types alternatif}\label{sec:mlalt}
Nous introduisons ici une manière de typer nos programmes {\em modulo}
isomorphismes de types, associant à un programme $e$ un type $\utau$,
où les variables de types ont été remplacées par la constante $\UTop$.


On définit le prédicat $\URule{\UGamma}{e}{\utau}$ par le système de
règles suivant:

\begin{mathpar}
\inferrule{ }{\URule{\UGamma}{n }{\TInt}}
\and
\inferrule{\utau \geninst \UGamma(p)}{\URule{\UGamma}{p}{\utau}}
\and
\inferrule{\utau \geninst \UGamma(r)}{\URule{\UGamma}{r}{\utau}}
\and
\inferrule{ \forall i=1..n, ~ \URule{\UGamma}{\e_i}{\utau_i}}
          {\URule{\UGamma}{(\eunn) }{(\utau_1\times\ldots\times\tau_n)}}
\and
\inferrule{\C{i}\in\Delta(\TConstr{T}{\vector\utau})}
          {\URule{\UGamma}{\C{i} }{T(\vector\utau)}}
\and
\inferrule{\F{i}{\utau'}\in\Delta(\TConstr{T}{\vector\utau})
  \\ {\URule{\UGamma }{ e }{ \utau'}}}
          {\URule{\UGamma}{\F{i}{\e} }{T(\vector\utau)}}
\and
\inferrule{{\URule{\UGamma }{ e' }{ (\utau_1\times\ldots\times\utau_n)}} \\ 
    {\URule{\UGamma \oplus \{ p_i: \utau_i
 \}_{i=1..n}}{\e}{\utau}}}
          {\URule{\UGamma}{\Let{\punn}{\e'}{\e} }{\utau}}
\and
\inferrule{
    \URule{\UGamma \oplus \{ r_i: \utau'_i \}_{i=1..n}}{\e}{(\utau'_1\times \ldots\times \utau'_n)}}
          {\URule{\UGamma}{\Fix{\runn}{e} }{ (\utau_1\times\ldots\times\utau_n)}}
\quad (\forall i=1..n,\; \utau_i \geninst {\utau'_i})
\end{mathpar}

\begin{Lemme}
Si $\UGamma' \geninstr \UGamma$ et 
\inferrule{}{\URule{\UGamma}{e }{\utau}}, alors 
\inferrule{}{\URule{\UGamma'}{e }{\utau}}.
\end{Lemme}
\begin{SPreuve}
Par induction sur l'arbre de typage de \inferrule{}{\URule{\UGamma}{e}{\utau}}.
\end{SPreuve}

\subsection{Équivalence entre  $\TrecRulez$ et $\URulez$}

\begin{Lemme}\label{lemme:equivalence-ml-u}
Soient un programme $\e$, un type $\tau_e$  et
un environnement $\Gamma$. 
Si \inferrule{}{\TrecRule{ \Gamma}{\e}{\tau_e }} 
alors pour $\UGamma = \univ\Gamma$ et $\utau_e = \univ{\tau_e}$ on a:
\inferrule{}{\URule{ \UGamma}{\e}{\utau_e}}.
\end{Lemme}

\begin{SPreuve}
Par réécriture systématique de l'arbre de typage de  $\TrecRule{
  \Gamma}{\e}{\tau_e}$.
\end{SPreuve}

\begin{Theoreme}\label{theoreme:equivalence-u-ml}
{\bf (Équivalence)}
Soient un programme $\e$, un type $\utau$  et
un environnement $\UGamma$. 
Si \inferrule{}{\URule{\UGamma}{\e}{\utau}} alors 
pour $\Gamma = \Tschema{\UGamma}$ 
et pour tout $\tau \inst \Tschema{\utau}$ on a:
\inferrule{}{\TrecRule{\Gamma}{\e}{\tau}}.
\end{Theoreme}

\begin{SPreuve}
Par induction sur la dérivation du jugement
\inferrule{}{\URule{\UGamma}{e}{\utau}} et par cas selon la dernière
règle utilisée.
\end{SPreuve}

\subsection{Correction}

\begin{Lemme}
\label{lemme:check-correction-gamma-structure}
Soient une valeur $\v$, un type $\utau$, 
et un environnement de type $\UGamma$ tels que 
$\fv{\v} \subseteq \dom{\UGamma}$.
Si $\check{\UGamma}{\utau}{\v} = (e,\UGamma')$ 
alors:
\begin{enumerate}
\item \label{struct:domain} $\dom{\UGamma'} = \dom{\UGamma}$,
\item \label{struct:instance} $\UGamma' \geninstr \UGamma$,
\item \label{struct:serial} $\serial{\e}=\v$.
\end{enumerate}
\end{Lemme}

\begin{SPreuve}
Par induction sur les appels récursifs de $\check{\UGamma}{\utau}{\v}$.
\end{SPreuve}

\begin{Lemme}
\label{lemme:check-correction-induction}
Soient une valeur $\v$, un type $\utau$, 
et un environnement de type $\UGamma$ 
tels que $\fv{\v} \subseteq \dom{\UGamma}$.
Si $\check{\UGamma}{\utau}{\v} = (e, \UGamma')$  
alors \inferrule{}{\URule{\UGamma'}{e }{\utau}}.
\end{Lemme}

\begin{SPreuve}
Par induction sur les appels récursifs de $\check{\UGamma}{\utau}{\v}$.
\end{SPreuve}

\begin{Theoreme}\label{theoreme:check-correction}
{\bf (Correction)}
Soient une valeur $\v$, et un type $\utau$.
Si $\check{\emptyset}{\utau}{\v}$ réussit en calculant un
programme $e$, alors:
\inferrule{}{\URule{\emptyset}{e }{\utau}}.
\end{Theoreme}

\begin{Preuve}
Le lemme \ref{lemme:check-correction-gamma-structure} nous indique
 $\check{\emptyset}{\utau}{\v} = (e,\emptyset)$.
 C'est ensuite une application directe du
lemme \ref{lemme:check-correction-induction}.
\end{Preuve}
Le théorème \ref{theoreme:check-correction} et le lemme
\ref{lemme:check-correction-induction} nous indiquent que si la
vérification de $v$ réussit en $\utau$, alors l'algorithme produit un
programme $e$ dont la valeur est $v$ et qui est typable par
$\utau$. Cela rend bien évidemment inutile la production de $e$ dans
une implémentation effective de l'algorithme, qui peut se borner à
rendre $v$ en résultat en cas de succès.

\subsection{Complétude}

Notre algorithme n'est pas complet dans le sens où toute expression
admettant une dérivation de typage dans $\TrecRulez$ ne sera pas 
forcément acceptée par notre algorithme.
En effet, les définitions de types non réguliers fournissent des
contre-exemples, puisque certaines définitions de structures de
données récursives nécessitent un typage polymorphe de la récursion
pour être acceptées en ML. Soit par exemple la définition de type:
\begin{center}
$\TypeName{Nest}(\alpha) = \Tbty\ldots \Tor {\sf B} ({\TypeName{Nest}}
  (\alpha \times \alpha)) \Tor \ldots\Tety $
\end{center}
et considérons la valeur $\Fix{r}{{\sf B}(r)}$. On montre aisément que
cette valeur peut avoir le type ${\TypeName{Nest}}({\TInt})$:
$$
\inferrule{
  \inferrule{
  \TrecRule{\{ r : \forall \alpha .{\TypeName{Nest}}(\alpha) \}}
           {r}{\TypeName{Nest}(\alpha \times\alpha)}}{
  \TrecRule{\{ r : \forall \alpha . {\TypeName{Nest}}(\alpha)\}}
           {{\sf B}\,(r)}{\TypeName{Nest}}(\alpha)}
}{
  \TrecRule
      {\emptyset}
      {\Fix{r}{{\sf B}\,(r)}}{\TypeName{Nest}(\TInt)}
}\qquad (\TypeName{Nest}(\TInt) \inst \forall \alpha . \TypeName{Nest}(\alpha))
$$
Cependant, lorsqu'on demande à l'algorithme de vérifier cette même
assertion, celui-ci va échouer en tentant de vérifier que
$\TypeName{Nest}(\TInt \times \TInt) \geninst \TypeName{Nest}(\TInt)$.

\medskip

Nous allons montrer par contre que notre algorithme est complet pour toute
valeur
typable sans avoir recours à la récursion polymorphe. Soit $\TRulez$
le système obtenu  de $\TrecRulez$ en y
remplaçant la règle de typage polymorphe de la récursion par la règle
classique de typage (monomorphe) de la récursion:
\begin{mathpar}
\inferrule{ 
    \TRule{\Gamma \oplus \{ r_i: \tau_i \}_{i=1..n}}{\e}{(\tau_1\times\ldots\times\tau_n)}}
          {\TRule{\Gamma}{\Fix{\runn}{\e}}{(\tau_1\times\ldots\times\tau_n)}}
\end{mathpar}
\begin{Lemme}
\label{lemme:check-complete-induction}
Soient une expression $e$, un type $\tau$ et un environnement $\Gamma$
tels que $\fv{e} \subseteq \dom{\Gamma}$. Soient une substitution
$\theta$ et un type $\utau$ tels que $\dom{\theta} = \fv{ \Gamma}
\cup \fv{\tau}$, et $\fv{\img{\theta}}\#\dom{\theta}$. Posons
$\utau = \univ{\theta(\tau)}$.

Si 
\inferrule{}{\TRule{\Gamma}{e}{\tau}} alors pour tout 
$\Gamma' \geninst \theta(\Gamma)$, on a
$\check{\univ{\Gamma'}}{\utau}{\serial{\e}} = (\e,\UGamma'')$ où
$\UGamma'' = \univ{\Gamma''}$ avec $\Gamma'' \geninst\theta(\Gamma)$.
\end{Lemme}
\begin{SPreuve}
Par induction sur la 
dérivation de typage
\inferrule{}{\TRule{\Gamma}{e}{\tau}}, et par cas selon la dernière
règle utilisée.
\end{SPreuve}

\begin{Theoreme}\label{theoreme:check-complete}
{\bf (Complétude)}
Soient une expression close $e$ et un type $\tau$.
Si \inferrule{}{\TRule{\emptyset}{e}{\tau}},
alors $\check{\emptyset}{\univ{\tau}}{\serial{e}}$ réussit.
\end{Theoreme}

\begin{Preuve}
C'est un cas particulier du lemme \ref{lemme:check-complete-induction}.
\end{Preuve}

\subsection{Complexité du parcours}
Pour une valeur sans cycle, l'algorithme de parcours est linéaire en la
taille $N$ du programme la représentant. En effet, le choix d'anti-unifier
le type de toutes les références à un bloc partagé avant de parcourir
ce dernier permet de ne parcourir qu'une seule fois l'ensemble de la valeur.
En présence de cycles il faut ajouter au coût de ce parcours celui de
l'identification et du tri des CFC, qui est dans le pire des
cas de l'ordre de $N\times P$ où $P$ est
la profondeur maximale d'imbrication de constructions $\Fixz$.

\section{Implantation}\label{sec:impl}

Une implantation prototype de cet
algorithme de vérification a été réalisée, ainsi que de certains éléments nécessaires
à son intégration en \ocaml{}. Cette section décrit quelques-uns des
points importants de cette mise en {\oe}uvre; le lecteur intéressé pourra se
référer à~\cite{henryMPRI} pour y trouver une description plus
précise.

\paragraph{Représentation des types}
Les types abstraits et la compilation séparée font que lorsqu'on
compile une valeur \ocamlcode{\trepr{$\tau$}}, on ne dispose pas
statiquement de toutes les informations nécessaires à la
reconnaissance des valeurs de type $\tau$.  On résoud ce problème en
compilant chaque déclaration de type comme une fonction ayant autant
de paramètres que le constructeur de types. Ainsi, dans une expression
\ocamlcode{\trepr{$\tau$}}, une référence à un type abstrait sera
compilée comme un appel à la fonction correspondante dont le code ne sera accessible que lors de l'exécution, puisque fourni
par l'implémentation de l'interface exportant
ce type. Cette solution présente l'avantage de laisser inchangées les
signatures de modules et de ne modifier que les implémentations.

\paragraph{Détection du partage et tri topologique}
Nous utilisons pour la recherche des CFC et leurs tris topologiques
l'algorithme proposé par Tarjan \cite{Tarjan1972}. Les informations
sur le partage peuvent être accessibles en temps constant en
encapsulant chaque valeur partagée dans un bloc spécial lors de la
reconstruction en mémoire de la valeur dé-sérialisée. Cette capsule
permet d'accéder aux informations nécessaires à l'algorithme de
recherche de CFC, ainsi qu'au résultat de l'anti-unification des
contraintes de type déjà rencontrées pour le bloc
encapsulé. Chaque référence sur cette capsule n'étant suivie qu'une
seule fois, le graphe mémoire réel de la valeur peut être reconstitué
incrémentalement en remplaçant systématiquement la référence sur la
capsule par une référence sur le bloc encapsulé lors du parcours de
vérification.

L'algorithme de parcours d'une valeur, présenté à la
section~\ref{sec:informel}, parcourt les blocs du graphe mémoire d'une
valeur dans le même ordre que l'algorithme $\checkz$. Chaque bloc
partagé introduit une nouvelle construction $\Let{p}{\e}{\f}$ où $p$
est une variable fraîche, $\f$ le contexte déjà exploré et $\e$ la
valeur représentée par ce bloc. Chaque CFC est représentée par une
construction $\Fixz$ introduisant une variable pour chacune de ses
racines. Des constructions $\Fixz$ imbriquées représentent des CFC
internes à une CFC.

\section{Travaux connexes}\label{sec:connexes}


Les solutions apportées jusqu'à maintenant à ce problème se résument à
sérialiser non seulement une valeur, mais aussi son type sous une
forme ou sous une autre. Au début des années 1980, Herlihy et Liskov
ont proposé une méthode de transmission de valeurs s'appliquant aux
types abstraits dans le langage CLU~\cite{CLU}: les valeurs sont
sérialisées avec leur type et des fonctions d'encodage et de décodage
spécifiques fournies par les types abstraits sont utilisées à la place
des mécanismes généraux de sérialisation.  Leroy et
Mauny~\cite{leroy93dynamics} ont étudié l'introduction de valeurs à
types dynamiques en ML, étendant ainsi une proposition initiale de
Cardelli dans le langage Amber~\cite{DBLP:conf/litp/Cardelli85}. Les
valeurs à types dynamiques (ou plus simplement {\em dynamiques}) sont
des paires composées d'une valeur $v$ et d'un type $\tau$ telles que
$v:\tau$. La création de dynamiques nécessite la collaboration du
compilateur, qui inclut le type statique $\tau$ de la valeur $v$ afin
de créer le dynamique $(v,\tau)$. Pour passer du dynamique $(v,\tau)$
à la valeur typée $v:\tau$ (le «déconstruire»), il est nécessaire
d'avoir recours à un mécanisme particulier de filtrage qui réalise des
tests de type sur les dynamiques.  Puisque les dynamiques ont tous le
même type, ils sont à même d'être manipulés par des fonctions de
lecture et d'écriture typées.  Les fonctions de sérialisation et de
dé-sérialisation ont alors respectivement les types \ocamlcode{Dynamic
$\to$ Filename $\to$ Unit} et \ocamlcode{Filename $\to$ Dynamic}. On
se retrouve alors dans une situation similaire au mécanisme de
sérialisation actuellement implanté en Java.

Les travaux de Dubois, Rouaix et Weis sur le polymorphisme générique
\cite{DBLP:conf/popl/DuboisRW95} ont permis à Furuse et Weis de
concevoir une forme de valeurs à types dynamiques, et de proposer des
fonctions de lecture et d'écriture de valeurs avec leurs
types. L'information de type qui est écrite est une empreinte
cryptographique du type d'origine de la valeur, {\em modulo} renommage de
labels et de constructeurs, fournissant ainsi une certaine souplesse
grâce à la possibilité de renommage, et une efficacité des tests de
type --- nécessaires à la déconstruction des dynamiques --- puisque
seules des empreintes doivent être comparées.

Leifer, Peskine, Sewell et Wansbrough~\cite{Leifer2003} s'intéressent
quant à eux à la préservation de l'abstraction, renonçant donc à la
souplesse mentionnée plus haut. Pour garantir la préservation de
l'abstraction, ils associent aux valeurs l'empreinte cryptographique
des définitions de leur type et de leur contexte (c'est-à-dire
essentiellement le texte du module les contenant ainsi que les modules
qui y sont importés).

Du côté du monde objet à la Java, la dé-sérialisation d'un objet
produit une instance du type {\tt{Object}}.  C'est ensuite le
mécanisme de {\ang{downcasting}} qui en assurera la spécialisation à
la demande.

L'approche que nous suivons se différencie clairement de ces travaux,
en ce que, d'une part, nous considérons la dé-sérialisation de valeurs qui ne
portent aucune information de type, et d'autre part, la vérification
de typage que nous effectuons allie garantie et souplesse.

\bigskip

Le problème de la reconstruction dynamique du type des valeurs dans
les langages statiquement typés comme ML a surtout été étudié dans le
cadre de l'interaction entre l'optimisation de la représentation des
données et la gestion automatique de la mémoire (voir par exemple
\cite{DBLP:journals/lisp/Appel89, DBLP:conf/lfp/GoldbergG92,
  aditya93compilerdirected, tolmach94tagfree,
chailloux-types}), ainsi que dans le but de concevoir des outils de
mise au point. La difficulté majeure à surmonter dans ce contexte est
de retrouver les {\em instances} d'utilisation de valeurs
polymorphes. En effet, un gestionnaire mémoire ou un {\em debugger}
est amené à manipuler des valeurs résultant de calculs qui ont
impliqué des utilisations d'instances particulières de valeurs
polymorphes, et la reconstruction de ces
instanciations est essentielle à l'obtention d'informations de types
fiables sur la mémoire. Par exemple, une liste paramètre de la fonction
\ocamlcode{List.length} a pour type statique
\ocamlcode{List($\alpha$)} alors qu'un
argument effectif de cette même fonction peut être de type
\ocamlcode{List(Int)} ou bien \ocamlcode{List(Bool$\times$Float)}.  Le
problème de la reconstruction dynamique de types pour la gestion de la
mémoire ou la conception d'outils de mise au point consiste donc à
retrouver cette instance, en reconstituant l'histoire du calcul.

En règle générale, cette reconstruction impose au programme de
stocker suffisamment d'informations de type à l'exécution pour pouvoir
reconstruire cet historique, pour garder trace des applications de
fonctions polymorphes et prendre en compte les mutations et les levées
d'exceptions. Par contre, ces travaux font l'hypothèse que cette
information de type est correcte: on reconstruit donc le type d'un
programme à un moment de son exécution mais on ne procède à aucune
vérification.

Le travail présenté ici ne nécessite pas de reconstruire
ainsi le type d'un programme, mais vise au contraire à {\em vérifier}
qu'une valeur appartient bien à un type.

\section{Discussion et travaux futurs}\label{sec:futur}

Un traitement correct des données mutables est indispensable avant
d'envisager toute intégration dans un langage de programmation. Or,
une utilisation na\"ive de cet algorithme de vérification sur des
types de données mutables peut être source d'incohérences, lorsque le
partage d'une valeur mutable lui impose d'avoir un type polymorphe,
produit par anti-unification. Une solution évidente serait
d'interdire de telles anti-unifications en faisant échouer la
vérification lorsque celle-ci est amenée à anti-unifier deux termes
partageant un même symbole de tête (constructeur d'un type mutable)
dont les arguments ne sont pas identiques.  Par exemple, si
{\ocamlcode Ref} est un type paramétré dont les valeurs sont des
enregistrements avec un champ mutable, et que l'algorithme est amené à
anti-unifier {\ocamlcode Ref($\utau_1$)} et {\ocamlcode
Ref($\utau_2$)}, il échouera si $\utau_1$ et $\utau_2$ ne sont pas
identiques.  De tels cas d'échec peuvent être vus comme le refus du
partage potentiel d'une occurrence mutable vue par le contexte comme
porteuse de valeurs de types incompatibles: comme dans le cas non
mutable, cela correspond à une absence de valeur au moment de la
dé-sérialisation, mais le caractère mutable ne garantit pas la
pérennité de cette absence.

Nous n'avons pas abordé le problème des valeurs fonctionnelles dans
cet article. \ocaml{} permet leur sérialisation, sous la forme,
essentiellement, de fermetures composées d'une adresse de code et d'un
environnement. La de-sérialisation est capable de relocaliser de façon
sûre l'adresse de code; il reste donc, pour une dé-sérialisation sûre,
à obtenir et vérifier le type de l'environnement. Or, ce dernier n'est
généralement pas disponible: en effet le type d'une valeur
fonctionnelle n'apporte en général pas d'information sur le type de
son environnement. Nous envisageons dans un futur proche d'aborder ce
problème.


Le cas des valeurs de type \ocamlcode{TRepr($\tau$)} mérite une
attention particulière: en effet, elles peuvent elles aussi être
sérialisées, et devront être dé-sérialisées de façon sûre.  Par
exemple, la sérialisation de la valeur
\ocamlcode{«$\forall\alpha$.List($\alpha$)»} doit pouvoir être relue
comme étant de type \ocamlcode{TRepr(List(Int))} ou
\ocamlcode{$\forall\alpha$.TRepr(List($\alpha$))}, mais pas
\ocamlcode{$\forall\alpha$.TRepr($\alpha$)}.  Il apparaît donc
souhaitable d'autoriser la lecture de la représentation d'un type
comme étant du type de la représentation d'une de ses instances.

Le type \ocamlcode{TRepr} étant prédéfini, il est aisé d'étendre
l'algorithme $\checkz$ par une analyse explicite des différents cas
constituant le corps de sa définition, et faisant en sorte qu'il
accepte la représentation de $\top$ comme étant du type de la
représentation d'un type quelconque.

\section{Conclusion}
Nous avons présenté dans cet article une méthode originale de
vérification de types de valeurs dé-sérialisées, et nous en avons
prouvé la correction. Une implantation prototype en a démontré
l'effectivité, et nous envisageons l'extension de la méthode aux
données mutables et fonctionnelles. 

\section*{Remerciements}
Nous tenons à remercier Damien Doligez et Didier Rémy pour les
discussions que nous avons pu avoir avec eux à différentes étapes de
ce travail, ainsi que les rapporteurs dont les commentaires nous ont
incité à clarifier certains passages de cet article.

\bibliography{main}

\begin{thebibliography}{10}

\bibitem{aditya93compilerdirected}
S.~\bgroup\sc Aditya\egroup{} \andname{} A.~\bgroup\sc Caro\egroup{}.
\newblock << Compiler-directed Type Reconstruction for Polymorphic Languages
  >>.
\newblock \Inname{} {\em Proceedings of the ACM Conference on Functional
  Programming Languages and Computer Architecture}, \pagesname{} 74--82, 1993.

\bibitem{DBLP:journals/lisp/Appel89}
A.~W. \bgroup\sc Appel\egroup{}.
\newblock << Runtime Tags Aren't Necessary >>.
\newblock 2(2):153--162, 1989.

\bibitem{DBLP:conf/litp/Cardelli85}
L.~\bgroup\sc Cardelli\egroup{}.
\newblock << Amber >>.
\newblock \Inname{} G.~\bgroup\sc Cousineau\egroup{}, P.-L. \bgroup\sc
  Curien\egroup{} \andname{} B.~\bgroup\sc Robinet\egroup{}, \editornames{},
  {\em Combinators and Functional Programming Languages}, \volumename{} 242
  \ofname{} {\em Lecture Notes in Computer Science}, \pagesname{} 21--47.
  Springer, 1985.

\bibitem{chailloux-types}
E.~\bgroup\sc Chailloux\egroup{}, P.~\bgroup\sc Manoury\egroup{} \andname{}
  B.~\bgroup\sc Pagano\egroup{}.
\newblock << Types behind the mirror: a proposal for partial ML type
  reconstruction at runtime >>.
\newblock \Inname{} {\em Workshop on Types in Compilation}, 1997.

\bibitem{crary98intensional}
K.~\bgroup\sc Crary\egroup{}, S.~\bgroup\sc Weirich\egroup{} \andname{}
  G.~\bgroup\sc Morrisett\egroup{}.
\newblock << Intensional Polymorphism in Type-Erasure Semantics >>.
\newblock \Inname{} {\em International Conference on Functional Programming},
  \pagesname{} 301--312, 1998.

\bibitem{DBLP:journals/jfp/Cosmo93}
R.~\bgroup\sc {Di Cosmo}\egroup{}.
\newblock << Deciding Type Isomorphisms in a Type-Assignment Framework >>.
\newblock {\em Journal of Functional Programming}, 3(4):485--525, 1993.

\bibitem{DBLP:conf/popl/DuboisRW95}
C.~\bgroup\sc Dubois\egroup{}, F.~\bgroup\sc Rouaix\egroup{} \andname{}
  P.~\bgroup\sc Weis\egroup{}.
\newblock << Generic Polymorphism >>.
\newblock \Inname{} {\em Symposium on Principles of Programming Languages},
  \pagesname{} 118--129, 1995.

\bibitem{jfpwjfla2000}
J.~\bgroup\sc Furuse\egroup{} \andname{} P.~\bgroup\sc Weis\egroup{}.
\newblock << Entr\'ees-sorties de valeurs en {C}aml >>.
\newblock \Inname{} {\em Journ\'ees Francophones des Langages Applicatifs}.
  INRIA, 2000.

\bibitem{DBLP:conf/lfp/GoldbergG92}
B.~\bgroup\sc Goldberg\egroup{} \andname{} M.~\bgroup\sc Gloger\egroup{}.
\newblock << Polymorphic Type Reconstruction for Garbage Collection Without
  Tags >>.
\newblock \Inname{} {\em LISP and Functional Programming}, \pagesname{} 53--65,
  1992.

\bibitem{henryMPRI}
G.~\bgroup\sc Henry\egroup{}.
\newblock << Vers un typage sûr de la dé-sérialisation de valeurs {ML} >>.
\newblock Rapport de stage de {M}aster {MPRI}, Universit\'e Pierre et Marie
  Curie, 2005.
\newblock \url{http://mpri.master.univ-paris7.fr/stages-2005-rapports.html}.

\bibitem{HerlihyLiskov1982}
M.~P. \bgroup\sc Herlihy\egroup{} \andname{} B.~\bgroup\sc Liskov\egroup{}.
\newblock << A Value Transmission Method for Abstract Data Types >>.
\newblock {\em ACM Transactions on Programming Languages and Systems},
  4(4):527--551, 1982.

\bibitem{HicksTIC2000}
M.~\bgroup\sc Hicks\egroup{}, S.~\bgroup\sc Weirich\egroup{} \andname{}
  K.~\bgroup\sc Crary\egroup{}.
\newblock << Safe and flexible dynamic linking of native code >>.
\newblock \Inname{} R.~\bgroup\sc Harper\egroup{}, \editorname{}, {\em Workshop
  on Types in Compilation}, \volumename{} 2071 \ofname{} {\em Lecture Notes in
  Computer Science}, \pagesname{} 147--176. Springer-Verlag, 2000.

\bibitem{huet-these}
G.~\bgroup\sc Huet\egroup{}.
\newblock << {\em R\'esolution d'\'equations dans des langages d'ordre
  1,2,\ldots, $\omega$} >>.
\newblock Th\`ese d'\'etat, Universit\'e Paris 7, 1976.

\bibitem{kfoury-polyrec}
A.~J. \bgroup\sc Kfoury\egroup{}, J.~\bgroup\sc Tiuryn\egroup{} \andname{}
  P.~\bgroup\sc Urzyczyn\egroup{}.
\newblock << Type reconstruction in the presence of polymorphic recursion >>.
\newblock {\em ACM Transactions on Programming Languages and Systems},
  15(2):290--311, 1993.

\bibitem{Leifer2003}
J.~J. \bgroup\sc Leifer\egroup{}, G.~\bgroup\sc Peskine\egroup{}, P.~\bgroup\sc
  Sewell\egroup{} \andname{} K.~\bgroup\sc Wansbrough\egroup{}.
\newblock << Global abstraction-safe marshalling with hash types >>.
\newblock \Inname{} {\em International Conference of Functional Programming},
  2003.

\bibitem{ocaml}
X.~\bgroup\sc Leroy\egroup{}, D.~\bgroup\sc Doligez\egroup{}, J.~\bgroup\sc
  Garrigue\egroup{}, D.~\bgroup\sc Rémy\egroup{} \andname{} J.~\bgroup\sc
  Vouillon\egroup{}.
\newblock << {\em The Objective Caml system, release 3.08} >>.
\newblock INRIA-Rocquencourt, 2004.
\newblock \url{http://caml.inria.fr/pub/docs/manual-ocaml/}.

\bibitem{leroy93dynamics}
X.~\bgroup\sc Leroy\egroup{} \andname{} M.~\bgroup\sc Mauny\egroup{}.
\newblock << Dynamics in {ML} >>.
\newblock {\em Journal of Functional Programming}, 3(4):431--463, 1993.

\bibitem{CLU}
B.~H. \bgroup\sc Liskov\egroup{}, R.~A. \bgroup\sc Atkinson\egroup{},
  T.~\bgroup\sc Bloom\egroup{}, J.~E. \bgroup\sc Moss\egroup{}, J.~C.
  \bgroup\sc Schaffaert\egroup{}, R.~W. \bgroup\sc Scheifler\egroup{}
  \andname{} A.~\bgroup\sc Snyder\egroup{}.
\newblock {CLU} {R}eference {M}anual.
\newblock \Inname{} {\em Lecture Notes on Computer Science}. Springer-Verlag,
  1981.

\bibitem{webster}
\bgroup\sc {{M}erriam{-}{W}ebster}\egroup{}.
\newblock << {O}n{L}ine {D}ictionary >>.
\newblock \url{http://www.m-w.com/dictionary.htm}, 2007.

\bibitem{MycroftRec}
A.~\bgroup\sc Mycroft\egroup{}.
\newblock << Polymorphic Type Schemes and Recursive Definitions >>.
\newblock \Inname{} {\em Proceedings of the 6th Colloquium on International
  Symposium on Programming}, \pagesname{} 217--228, London, UK, 1984.
  Springer-Verlag.

\bibitem{plotkin-phd}
G.~\bgroup\sc Plotkin\egroup{}.
\newblock << {\em Automatic Methods of Inductive Inference} >>.
\newblock PhD thesis, University of Edinburgh, 1971.

\bibitem{JavaSerialization}
\bgroup\sc {Sun Microsystems}\egroup{}.
\newblock << Java Object Serialization Specification >>.
\newblock
  \url{http://java.sun.com/j2se/1.5.0/docs/guide/serialization/spec/serialTOC.%
html}, 2006.

\bibitem{Tarjan1972}
R.~E. \bgroup\sc Tarjan\egroup{}.
\newblock << Depth first search and linear graph algorithms >>.
\newblock {\em SIAM Journal on Computing}, 1(2):146--160, 1972.

\bibitem{tolmach94tagfree}
A.~P. \bgroup\sc Tolmach\egroup{}.
\newblock << Tag-Free Garbage Collection Using Explicit Type Parameters >>.
\newblock \Inname{} {\em Conference on Lisp and Functional Programming},
  \pagesname{} 1--11, 1994.

\end{thebibliography}
  
\thispagestyle{colloquetitle}
\end{document}